\documentstyle[11pt,paspconf,epsf]{article}

\begin{document}

\title{Distortion of the SED of High-z Luminous Infrared Galaxies by Strong Lensing}
\author{Klaus Martin Pontoppidan\altaffilmark{1} \& Tommy Wiklind\altaffilmark{2}}

\altaffiltext{1}{Astronomical Observatory, University of Copenhagen, Copenhagen, Denmark}
\altaffiltext{2}{Onsala Space Observatory, Sweden}

\begin{abstract}
We present a model to estimate the effect of differential magnification on the SED of high-z ULIRGs. It is found that 
the ratio of the high temperature component to the low temperature component can vary with up to a factor of ten with
source position on very small angular scales ($\sim 0.01''$). This means that a correction for differential magnification
is needed when deriving dust properties from strongly lensed sources.
\end{abstract}


This work was inspired by the discovery of high redshift galaxies with a very high IR luminosity emanating from
dust grains (Hughes 1998). It is expected that the thermal dust emission has a temperature gradient over a fairly 
large spatial region spanning from a temperature near the dust evaporation temperature ($T_{ev}\simeq 1500$K) to a much lower
background temperature of about $30$ K in a cautious analogy to the nearby ultraluminous infrared galaxies (ULIRGs). 
Partly because of selection effects a large percentage of the sources in the current sample are strongly lensed.
In the strong lensing region, there is a large spatial variation
of the magnification parameter, especially near caustics. It is therefore conceivable that thermal emission from 
a region in the source with a specific temperature is enhanced in comparison with thermal emission with different temperatures
from other regions (Blain 1999). The following presents a model which quantitatively tries to determine the importance of this
effect. 

The model consists of a gravitational potential modeling a single lens galaxy and a face-on disc-shaped source with a Gaussian
temperature distribution modeling a dusty torus surrounding an AGN. 
The lens is modeled by an elliptical potential.

In the more realistic case of a number of compact 
starbursts embedded in a clumpy torus, the temperature structure will be more complicated. For simplicity only the first case is studied. It is not 
difficult to include the second case in the model using a phenomenological approach to the temperature structure, but the 
qualitative conclusions will be the same.  

The dust emission is modeled in each point in the source by a modified blackbody spectrum, where
the power of the dust emissivity function, $\beta$, may vary between 1.5 and 2.0.
The calculation of the magnification is done with a straight forward rayshooting algorithm, which simply maps every point in the lens plane
to the source plane and sums the number of hits in each source pixel.

For the chosen temperature distribution the distortion appears as either an enhancement of a low temperature component compared to a high temperature 
component or vice versa. 
A natural distortion parameter, $\delta$, emerges from this, namely as the ratio between the high temperature enhancement and the low temperature 
enhancement.
A high value of $\delta$ means that the lensed SED looks hotter compared to the unlensed SED at a given source 
position. A low value for the distortion parameter on the other hand signifies a lower temperature. This resembles closely the inclination effects
seen in radiative transfer models of a dusty torus (Granato 1996). Consequently care should be taken when applying these models to lensed sources.  

\begin{figure}
\plottwo{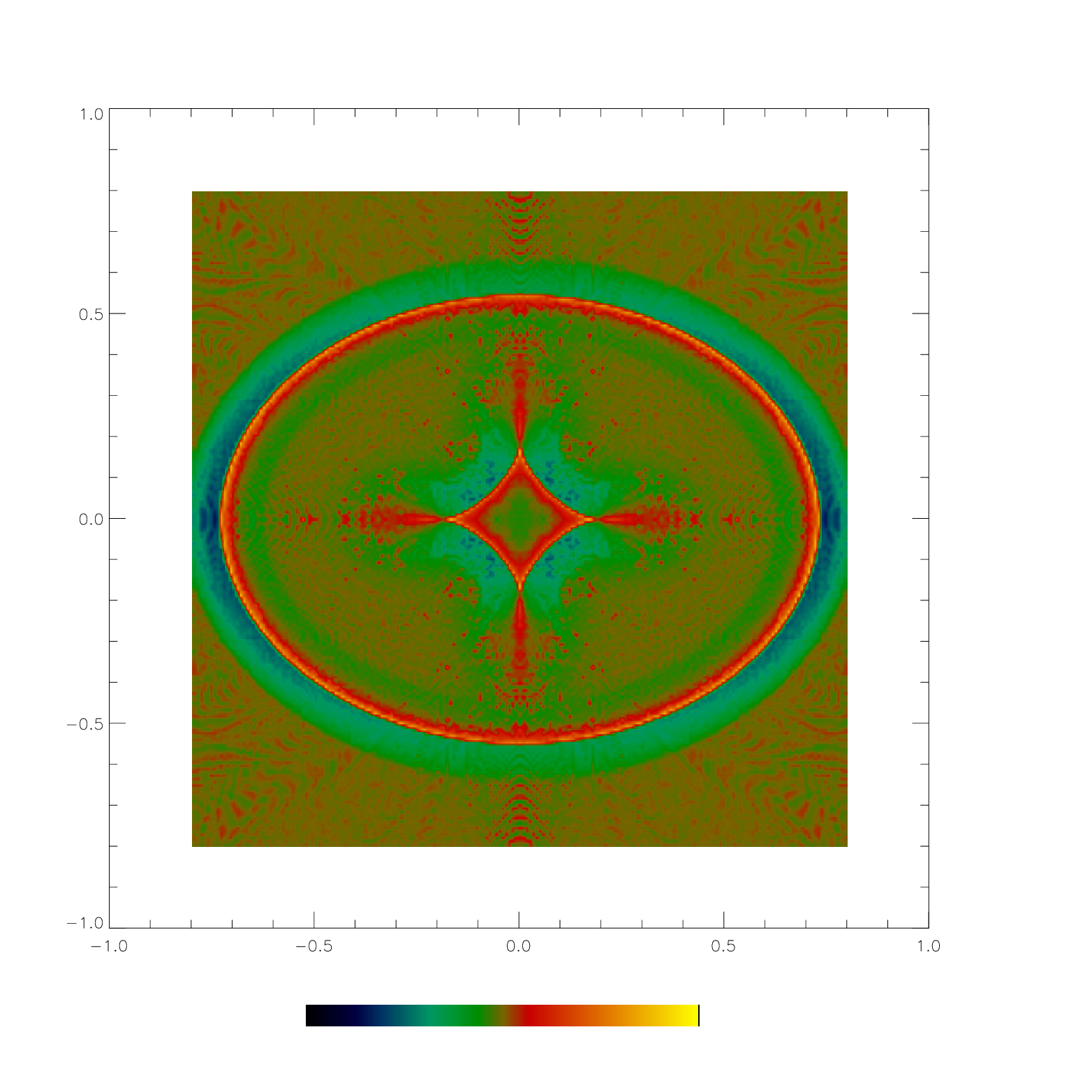}{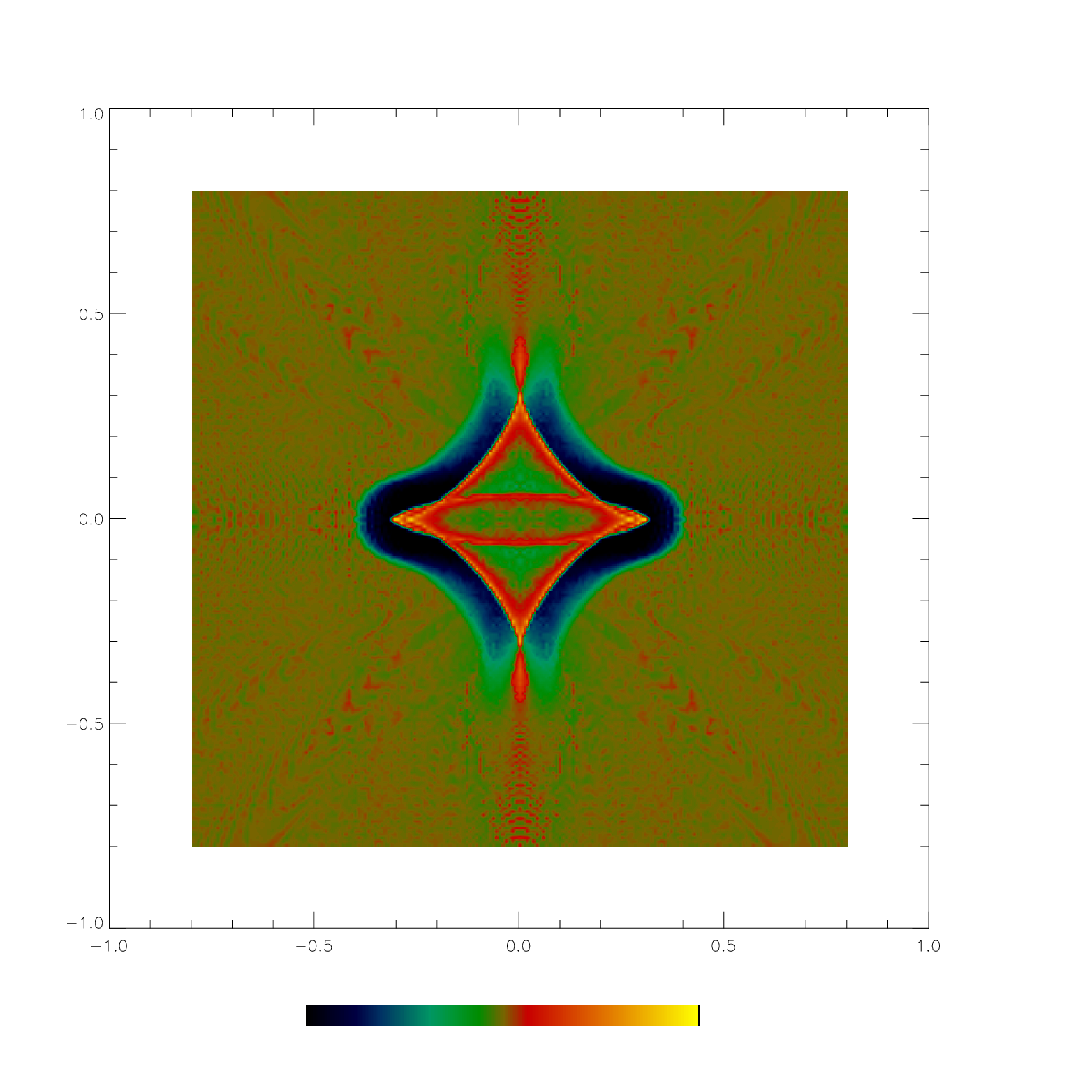}
\caption{This shows $\delta$ as a function of source position for two different models. The left is for a low ellipticity lens a 0.02'' source, while the 
right is for high lens ellipticity and a 0.04'' source.}
\label{dismap}
\end{figure}

Figure \ref{dismap} shows two distortion maps from the model.
The choice of parameters is fairly arbitrary but the qualitative results remain the same regardless of this choice.
The scale shown spans values of $\delta$ from -1 to 1, which means that in the most extreme cases, the ratio
between the two temperature components in the SED changes by an order of magnitude by the lensing. The structure of the distortion
near the caustics is very general.
The magnification parameter is everywhere larger inside a caustic than outside. So if a large
part of the source disc overlaps with the inner plateau while the the high temperature core lies outside, 
the low temperature part will be strongly enhanced. The maximum angular distance from the caustic of this effect is then 
simply determined by the angular size of the unlensed source. On the other hand, if the hot core of the source is made to 
coincide with a caustic, the core will be strongly enhanced compared to the low temperature disc. Thus
large variations in the distortion of the source SED over small angles are expected.

\end{document}